\documentclass[sigconf, nonacm]{acmart}
\newcommand\vldbdoi{}

\newcommand\vldbvolume{14}
\newcommand\vldbissue{1}

\newcommand\vldbavailabilityurl{}

\usepackage[noend]{algpseudocode}
\usepackage{algorithm}
\usepackage{bbm}
\usepackage{graphicx}
\usepackage{listings}
\usepackage[compress]{cleveref}
\usepackage{mathrsfs}
\usepackage{bm}
\usepackage{balance}
\usepackage{outlines}
\usepackage{svg}
\usepackage{grffile}
\usepackage[caption=false]{subfig}
\usepackage{xspace}
\usepackage{enumitem}
\usepackage{xcolor}
\usepackage{array}
\usepackage{nccmath}
\usepackage{multirow}
\usepackage{pifont}
\usepackage{soul}
\usepackage{caption}
\usepackage{upgreek}
\usepackage{bbm}
\usepackage{dsfont}
\usepackage{microtype}
\usepackage{amsfonts}
\usepackage{amsmath}
\usepackage{accents}
\usepackage{tabularx}
\usepackage[group-separator={,}]{siunitx}
\usepackage[bottom,hang,flushmargin]{footmisc}
\usepackage{setspace}
\usepackage{etoolbox}

\makeatletter
\preto{\@verbatim}{\topsep=0pt \partopsep=0pt }
\makeatother

\newcommand{\ubar}[1]{\underaccent{\bar}{#1}}

\newcommand{\lowoverline}[1]{%
  \overline{\smash{#1}\vphantom{x}}\vphantom{#1}%
}

\usepackage{xpatch}
\xpatchcmd{\NCC@ignorepar}{%
\abovedisplayskip\abovedisplayshortskip}
{%
\abovedisplayskip\abovedisplayshortskip%
\belowdisplayskip\belowdisplayshortskip}
{}{}

\graphicspath{./figures/}

\crefformat{section}{\S#2#1#3} %
\crefformat{subsection}{\S#2#1#3}
\crefformat{subsubsection}{\S#2#1#3}
\crefname{section}{Section}{Sections}
\Crefname{section}{Section}{Sections}
\crefname{figure}{Figure}{Figures}
\Crefname{figure}{Figure}{Figures}
\crefname{subfigure}{Figure}{Figures}
\Crefname{subfigure}{Figure}{Figures}
\crefname{algorithm}{Algorithm}{Algorithms}
\Crefname{algorithm}{Algorithm}{Algorithms}
\crefname{equation}{Equation}{Equation}
\Crefname{equation}{Equation}{Equation}
\crefname{lemma}{Lemma}{Lemma}
\Crefname{lemma}{Lemma}{Lemma}
\crefname{table}{Table}{Tables}
\Crefname{table}{Table}{Tables}
\crefrangelabelformat{subfigure}{#3#1#4--#5(\crefstripprefix{#1}{#2}#6}

\newcommand{\todo}[1]{\textcolor{red}{[TODO: #1]}}

\definecolor{emerald}{rgb}{0.31, 0.78, 0.47}

\newcommand{\revisioncolor}{black}
\newcommand{\revision}[1]{{\color{\revisioncolor} #1}}

\newcommand{\eat}[1]{}

\newcommand{\preeq}{\vspace{0mm}\begin{small}}
\newcommand{\posteq}{\vspace{0mm}\end{small}}

\newcommand{\system}{\textsc{MaskSearch}\xspace}

\usepackage{url}

\DeclareFixedFont{\ttb}{T1}{txtt}{bx}{n}{8}
\DeclareFixedFont{\ttm}{T1}{txtt}{m}{n}{8}

\usepackage{color}
\definecolor{deepblue}{rgb}{0,0,0.5}
\definecolor{deepred}{rgb}{0.6,0,0}
\definecolor{deepgreen}{rgb}{0,0.5,0}
\definecolor{purple}{rgb}{0.5,0,0.5}
\definecolor{gray}{rgb}{0.33,0.33,0.33}

\definecolor{dkgreen}{rgb}{0,0.6,0}
\definecolor{gray}{rgb}{0.5,0.5,0.5}
\definecolor{mauve}{rgb}{0.58,0,0.82}

\lstdefinelanguage{Python}{
	keywords={typeof, torch, nonzero, index_select, zeros_like, lt, masked_select, new, true, false, catch,def,val, function, return, null, catch, switch, var, shape,  while, do, else, case, break, override},
	keywordstyle=\color{blue}\bfseries,
	ndkeywords={class, export,extends, boolean, throw, implements, import, this, abstract, for, in, if},
	ndkeywordstyle=\color{dkgreen}\bfseries,
otherkeywords={+, =>,<=, ==, >,< , || , T},
	identifierstyle=\color{black},
	sensitive=false,
	comment=[l]{//},
	morecomment=[s]{/*}{*/},
	commentstyle=\color{purple}\ttfamily,
	stringstyle=\color{red}\ttfamily,
	morestring=[b]',
	morestring=[b]"
}
\begin{document}
\title{Demonstration of \system: Efficiently Querying Image Masks for Machine Learning Workflows}

\author{Lindsey Linxi Wei*}
\affiliation{%
  \institution{University of Washington}
}
\email{linxiwei@cs.washington.edu}

\author{Chung Yik Edward Yeung*}
\affiliation{%
  \institution{University of Washington}
}
\email{chungy04@cs.washington.edu}

\author{Hongjian Yu*}
\affiliation{%
  \institution{University of Washington}
}
\email{hjyu@cs.washington.edu}

\author{Jingchuan Zhou*}
\affiliation{%
  \institution{University of Washington}
}
\email{jzhou27@cs.washington.edu}

\author{Dong He}
\affiliation{%
  \institution{University of Washington}
}
\email{donghe@cs.washington.edu}

\author{Magdalena Balazinska}
\affiliation{%
  \institution{University of Washington}
}
\email{magda@cs.washington.edu}

\begin{abstract}
\begin{sloppypar}
We demonstrate \system, a system designed to accelerate queries over databases of image masks generated by machine learning models. 
\system formalizes and accelerates a new category of queries for retrieving images and their corresponding masks based on mask properties, which support various applications, from identifying spurious correlations learned by models to exploring discrepancies between model saliency and human attention. 
This demonstration makes the following contributions: (1) the introduction of \system's graphical user interface (GUI), which enables interactive exploration of image databases through mask properties, 
(2) hands-on opportunities for users to explore \system's capabilities and constraints within machine learning workflows, 
and (3) an opportunity for conference attendees to understand how \system accelerates queries over
image masks.

\end{sloppypar}
\end{abstract}

\maketitle

\begingroup
\renewcommand\thefootnote{}\footnote{\noindent*Equal contribution.

\noindent This work is licensed under the Creative Commons BY-NC-ND 4.0 International License. Visit \url{https://creativecommons.org/licenses/by-nc-nd/4.0/} to view a copy of this license. For any use beyond those covered by this license, obtain permission by emailing \href{mailto:info@vldb.org}{info@vldb.org}. Copyright is held by the owner/author(s). Publication rights licensed to the VLDB Endowment. \\
\raggedright Proceedings of the VLDB Endowment, Vol. \vldbvolume, No. \vldbissue\ %
ISSN 2150-8097. \\
\href{https://doi.org/10.48550/arXiv.2305.02375}{doi:\vldbdoi} \\
}\addtocounter{footnote}{-1}\endgroup

\ifdefempty{\vldbavailabilityurl}{}{
\vspace{.3cm}
\begingroup\small\noindent\raggedright\textbf{PVLDB Artifact Availability:}\\
The source code, data, and/or other artifacts have been made available at \url{\vldbavailabilityurl}.
\endgroup
}

\begin{sloppypar} %

\newcommand{\performanceFigure}{
    \begin{figure}[!t]
        \centering
        \includegraphics[width=0.85\linewidth]{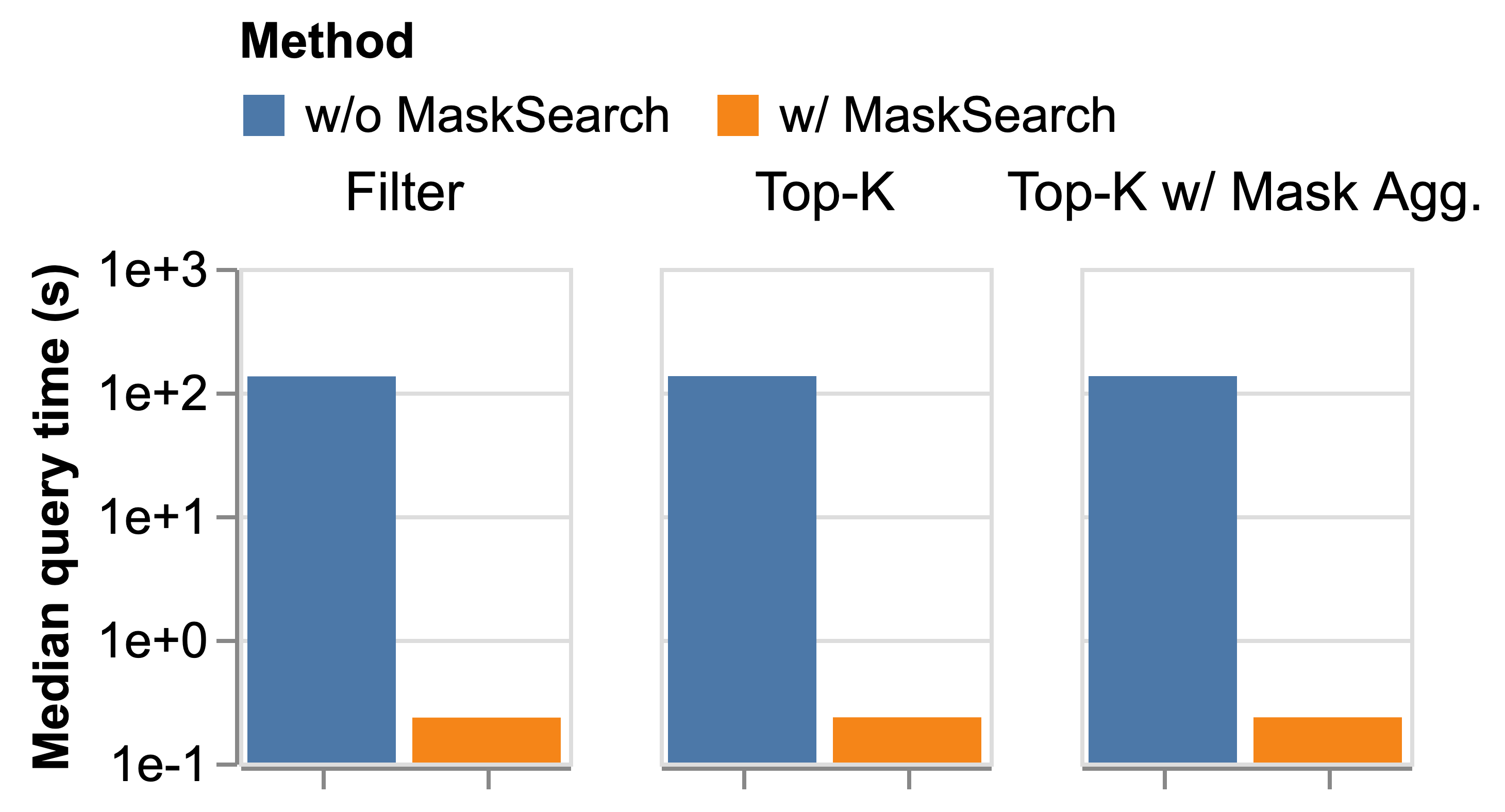}
        \vspace{-.5em}
        \caption{Comparison of median query times with and without \system for different query types.}
        \label{fig:performance}
    \end{figure}
}

\newcommand{\scenarioOneFigure}{
    \begin{figure*}[!t]
        \begin{center}
            \includegraphics[width=1.0\linewidth]{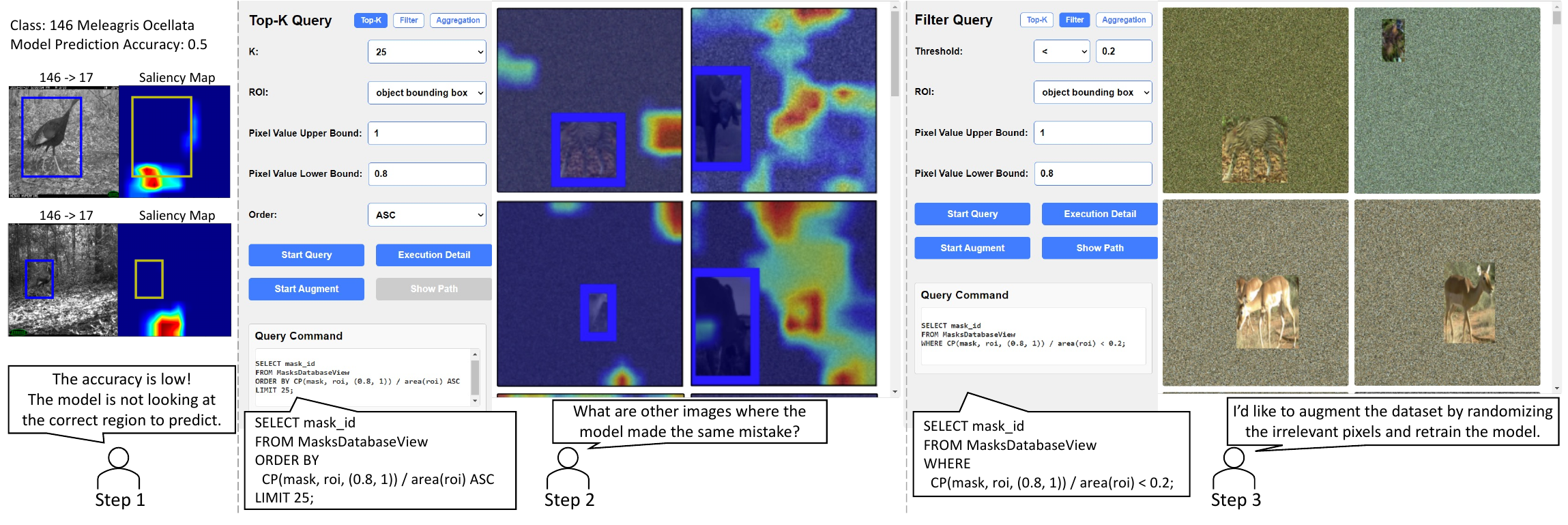}
        \end{center}
        \vspace{-1.5em}
        \caption{An example workflow of using \system's GUI in Scenario 1. In Step 1, 146 -> 17 means that the image with a ground truth label 146: Meleagris Ocellata was misclassified as class 17: Panthera Onca.}
        \label{fig:scenarioOne}
        \vspace{-1.0em}
    \end{figure*}
}

\newcommand{\scenarioTwoFigure}{
    \begin{figure}[!t]
        \centering
        \subfloat[before attack]{\includegraphics[width=3.5cm]{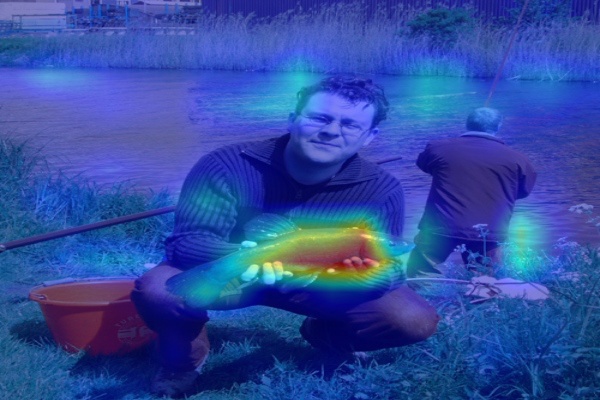}}
        \qquad
        \subfloat[after attack]{\includegraphics[width=3.5cm]{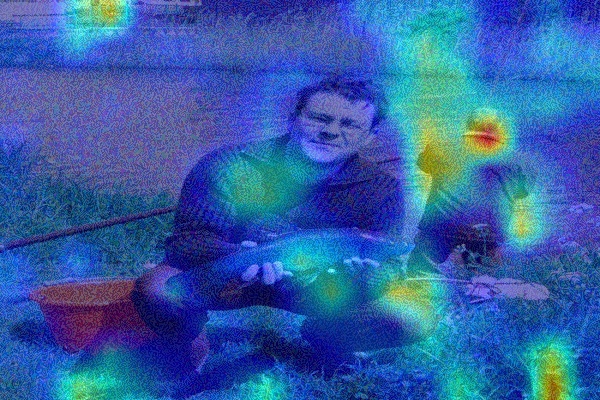}}
        \vspace{-1em}
        \caption{
        Saliency masks before and after a malicious attack on an example image from ImageNet~\cite{deng2009imagenet}. The object of interest in the image is the fish held by the man.}
        \label{fig:scenarioTwo}
        \vspace{-0.5em}
    \end{figure}
}

\newcommand{\inputFigure}{
    \begin{figure}[!t]
        \centering
        \includegraphics[width=0.99\linewidth]{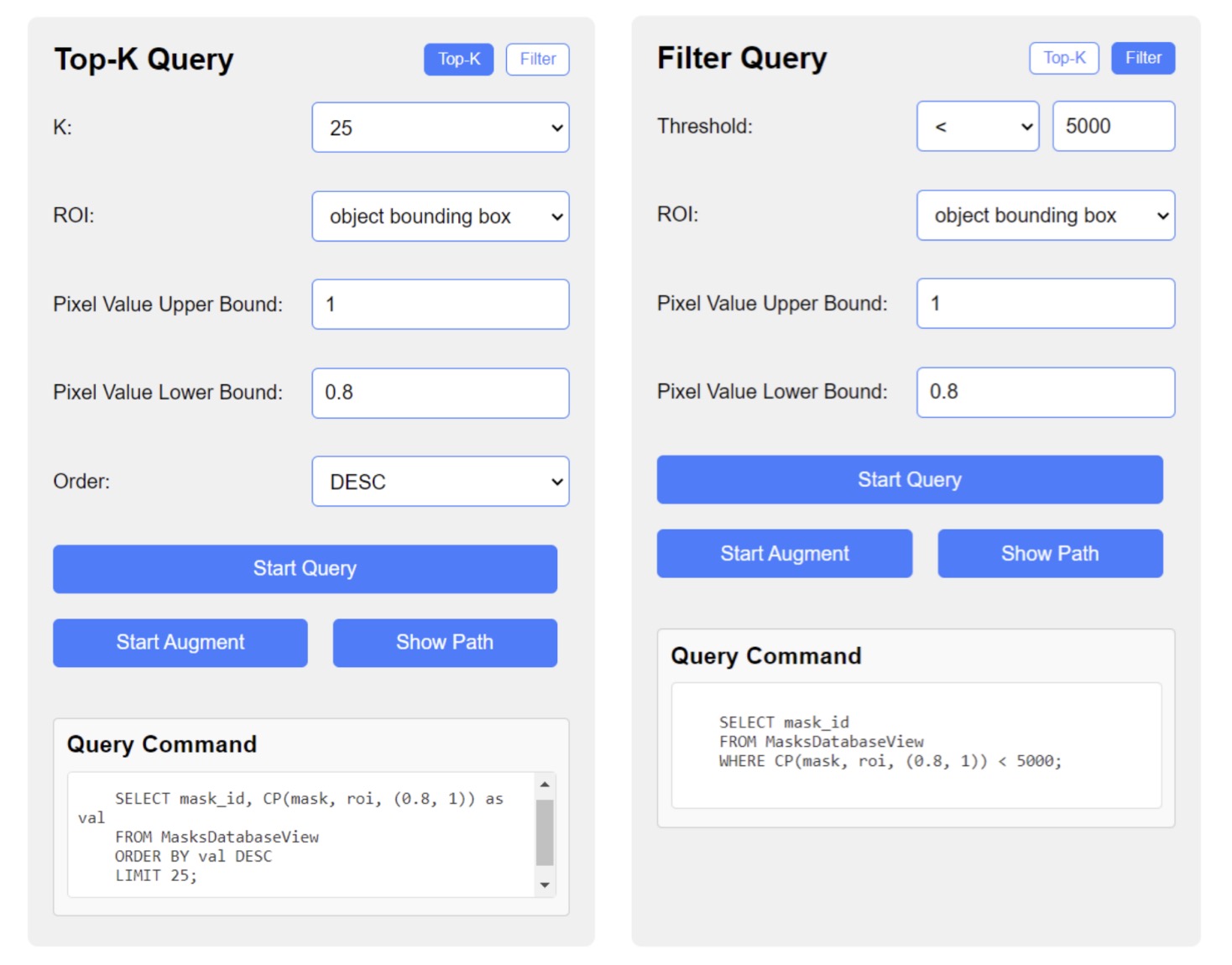}
        \caption{TBD.}
    \end{figure}
}

\newcommand{\explainMask}{
    \begin{figure}[!t]
        \centering
        \includegraphics[width=0.66\linewidth]{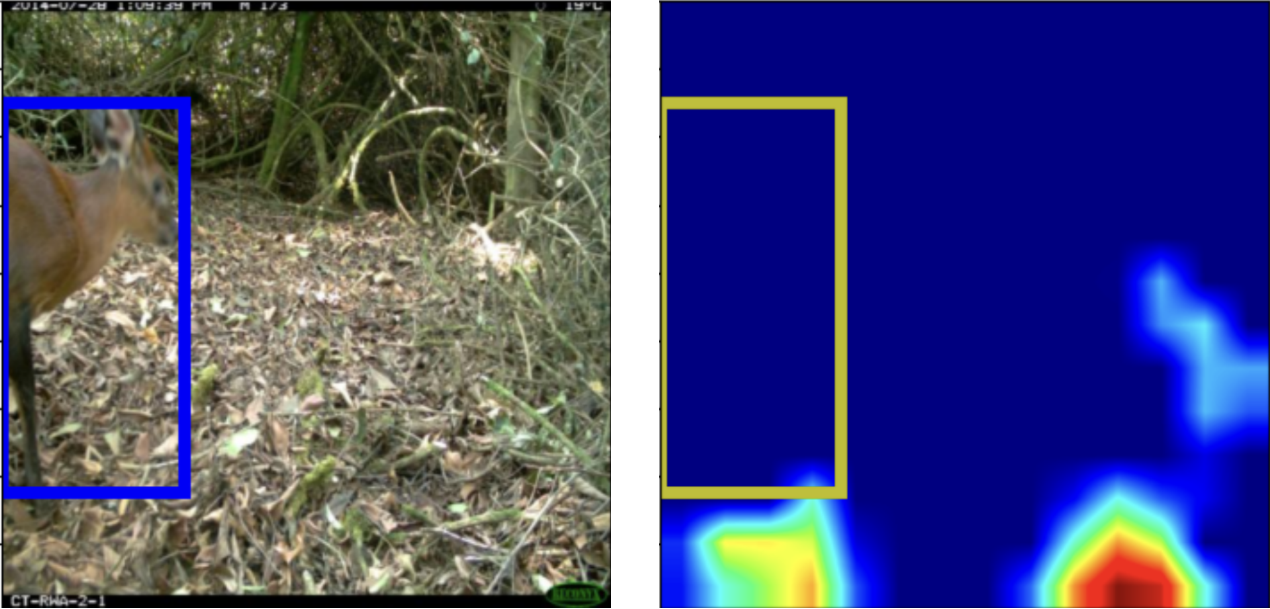}
        \vspace{-1em} 
        \caption{An example misclassified image and its model saliency map with the object bounding boxes (blue and yellow boxes). The salient pixels (red pixels in the saliency map) are focused on the background regions. This reveals that the model relies on irrelevant pixels to make the prediction.}
        \label{fig:explainMask}
        \vspace{-1em}
    \end{figure}
}

\newcommand{\HAFigure}{
    \begin{figure}[!t]
        \centering
        \includegraphics[width=0.99\linewidth]{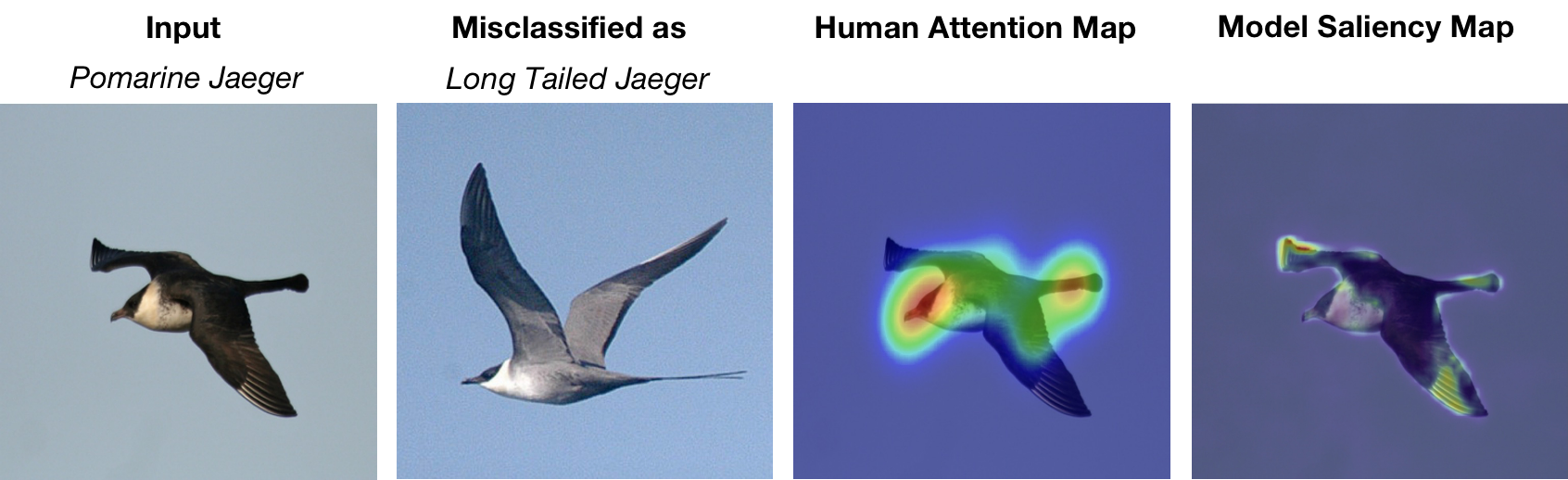}
        \vspace{-1em}
        \caption{Comparison of human attention maps and model saliency maps on images from CUB-200-2011~\cite{WahCUB_200_2011}. The human attention map shows that humans look at the head and tail of the Pomarine Jaeger to classify it, which are the discriminate traits. The model saliency map shows that the model is focusing on the wings instead. This explains why the model misclassifies Pomarine Jaeger as Long Tailed Jaeger.}
        \label{fig:HA}
        \vspace{-1em}
    \end{figure}
}

\newcommand{\maskExampleFigure}{
    \begin{figure}[t!]
        \begin{center}
            \subfloat[Segmentation mask]{\includegraphics[width=0.34\linewidth]{figures/image_segmentation_example.png}%
                \label{subfig:image-segmentation-example}}%
            \hfil
            \subfloat[Depth estimation mask]{\includegraphics[width=0.338\linewidth]{figures/depth_estimation_example.png}%
                \label{subfig:depth-estimation-example}}%
            \hfil
            \subfloat[Saliency map]{\includegraphics[width=0.25\linewidth]{figures/model_saliency_example.png}%
                \label{subfig:saliency-map-example}}%
        \end{center}
        \vspace{-1.4em}
        \caption{Examples of image masks that annotate image content for ImageNet~\cite{ILSVRC15} images produced by ML tasks.}
        \vspace{-1.5em}
        \label{fig:mask-example}
    \end{figure}
}

\newcommand{\introExampleFigure}{
    \begin{figure}[t!]
        \centering
        \includegraphics[width=0.85\linewidth]{figures/intro_example.pdf}
        \vspace{-1.5em}
        \caption{
        Example image masks: ImageNet~\cite{ILSVRC15} images overlaied with saliency maps. 
        Saliency maps in columns b) and c) reveal that the models rely on irrelevant pixels to make predictions. 
        Retrieving more examples with similar mask properties helps to better investigate the model's behavior. 
        }
        \vspace{-1.8em}
        \label{fig:intro-example}
    \end{figure}
}

\newcommand{\maliciousAttackFigure}{
    \begin{figure}[t!]
        \centering
        \includegraphics[width=\linewidth]{figures/malicious_attack_example.png}
        \vspace{-1.2em}
        \caption{\todo{TBD. Also need to include high-level queries that \system helps to support. And perhaps only show one row of attacked images to save space}.}
        \vspace{-0.53em}
        \label{fig:malicious-attack}
    \end{figure}
}

\newcommand{\queryExampleFigure}{
    \begin{figure}[t!]
        \begin{center}

            \subfloat[Example image]{\includegraphics[width=0.25\linewidth]{figures/query-example-image.png}%
                \label{subfig:query-example-image}}%
            \hfil
            \subfloat[Mask]{\includegraphics[width=0.25\linewidth]{figures/query-example-mask.png}%
                \label{subfig:query-example-mask}}%
        \end{center}
        \vspace{-1.2em}
        \caption{A toy image motivated by~\cite{degrave2021ai} and its mask. The purple box is the ROI. Predicates on masks often involve counting the number of pixels in the ROI with values in a range, e.g., $\#$ pixels in the ROI with values in $(0.85, 1.0)$ is $2$.}
        \vspace{-1.0em}
        \label{fig:query-example}
    \end{figure}
}

\newcommand{\notationTable}{
    \begingroup
    \setlength{\tabcolsep}{2pt}
    \begin{table}[t!]
        \small
        \centering
        \caption{Summary of frequently used notation.}
        \vspace{-1.1em}
        \begin{tabular}{ l l }
            \toprule
            Symbol & Meaning \\
            \midrule
            $\texttt{CP}(mask, r, (lv, uv))$ & Count of pixels in region $r$ of $mask$ \\
             & with pixel values in range $(lv, uv)$ \\
            $P$ & Predicate $\texttt{CP}(mask, roi, (lv, uv)) > T$ \\
            $\theta$ & Actual value of $\texttt{CP}(mask, roi, (lv, uv))$ \\
            $\bar{\theta}$ & Upper bound of $\theta$ \\
            $\ubar{\theta}$ & Lower bound of $\theta$ \\
            $\Delta$ & Size of a pixel value bin \\
            $C(mask\_id, r)$ & Histogram of reverse cumulative pixel counts \\
            $C(mask\_id, r)[i]$ & $\texttt{CP}(mask, r, (p_{min} + i\Delta, p\_max))$ \\
            $roi$ & Region of interest specified by the user \\
            $\lowoverline{roi}$ & Smallest region \textit{available} in CHI covering $roi$ \\
            $\underline{roi}$ & Largest region \textit{available} in CHI covered by $roi$ \\
            \bottomrule
        \end{tabular}
        \normalsize
        \label{tab:notation}
    \end{table}
    \endgroup
}

\newcommand{\chiIllustrationFigure}{
    \begin{figure}
    \centering
    \includegraphics[width=\linewidth]{figures/chi_illustration.png}
    \vspace{-2.0em}
    \caption{An example of CHI, \texttt{CP}, \textit{available region}, and $C$.}
    \vspace{-1.0em}
    \label{fig:chi-illustration}
    \end{figure}
}

\newcommand{\upperBoundIllustrationFigure}{
    \begin{figure}
    \centering
    \includegraphics[width=0.85\linewidth]{figures/upper_bound_illustration.png}
    \vspace{-1.2em}
    \caption{An example of \system computing the upper bounds, $\bar{\theta}_1$ and $\bar{\theta}_2$, given a mask, $roi$, and $(lv, uv)$.}
    \vspace{-1.0em}
    \label{fig:upper-bound-illustration}
    \end{figure}
}

\newcommand{\additiveFunctionIllustrationFigure}{
    \begin{figure}
    \centering
    \includegraphics[width=0.85\linewidth]{figures/additive_function_illustration.png}
    \vspace{-1.0em}
    \caption{Illustration of $\texttt{CP}$ being a (finitely)-additive function.}
    \vspace{-1.0em}
    \label{fig:additive-function-illustration}
    \end{figure}
}

\newcommand{\queryBasedOnMotivationTable}{
    \begingroup
    \setlength{\tabcolsep}{2pt}
    \begin{table*}[t!]
        \footnotesize
        \centering
        \caption{Summary of evaluated queries based on motivation and related work.}
        \vspace{-1.1em}
        \begin{tabularx}{\textwidth}{ l X }
            \toprule
            Query & Description \\
            \midrule
            Q1 & Returns masks s.t. $\texttt{CP}(mask, roi, (lv, uv)) > \num{5000}$, $roi = ((50, 50), (200, 200))$, $(lv, uv) = (0.6, 1.0)$, $model\_id = 1$ \\
            Q2 & Returns masks s.t. $\texttt{CP}(mask, roi, (lv, uv)) > \num{15000}$, $roi = \text{object}$, $(lv, uv) = (0.8, 1.0)$, $model\_id = 1$\\
            Q3 & Returns top-25 masks with largest $\texttt{CP}(mask, roi, (lv, uv))$, $roi = ((50, 50), (200, 200))$, $(lv, uv) = (0.8, 1.0)$, $model\_id = 1$ \\
            Q4 & Returns top-25 images with largest $\texttt{mean}(\texttt{CP}(mask, roi, (lv, uv)))$ (groupby $image\_id$) for $mask$s associated with two models, $roi = \text{object}$, $(lv, uv) = (0.8, 1.0)$ \\
            Q5 & Returns top-25 images with largest $\texttt{CP}(\texttt{intersect}(mask), roi, (lv, uv))$ (groupby $image\_id$) for $mask$s associated with two models, $roi = \text{object}$, $(lv, uv) = (0.8, 1.0)$ \\
            \bottomrule
        \end{tabularx}
        \normalsize
        \vspace{-1.0em}
        \label{tab:query-based-on-motivation}
    \end{table*}
    \endgroup
}

\newcommand{\masksLoadedTable}{
    \begingroup
    \setlength{\tabcolsep}{2pt}
    \begin{table}[t!]
        \footnotesize
        \centering
        \caption{Number of masks loaded during query execution. PG = PostgreSQL, TDB = TileDB, NP = NumPy.}
        \vspace{-1.1em}
        \begin{tabular}{ c c c c c c c c }
            \toprule
            Dataset & Method & Q1 & Q2 & Q3 & Q4 & Q5 \\
            \midrule
            \multirow{1}[2]{*}{\textit{WILDS}} & \system & 407 & 40 & 32 & 874 & 48 \\
             & PG \& TDB \& NP & \num{22275} & \num{22275} & \num{22275} & \num{44550} & \num{22275} \\
            \midrule
            \multirow{1}[2]{*}{\textit{ImageNet}} & \system & \num{2696} & \num{3849} & \num{2943} & \num{1494} & \num{2768} \\
             & PG \& TDB \& NP & \num{1331167} & \num{1331167} & \num{1331167} & \num{2662334} & \num{1331167} \\
            \bottomrule
        \end{tabular}
        \normalsize
        \vspace{-1.5em}
        \label{tab:masks-loaded}
    \end{table}
    \endgroup
}

\newcommand{\singleQueryBasedOnMotivationFigure}{
    \begin{figure*}[t!]
        \vspace{-2.3em}
        \begin{center}
            \captionsetup{font={color=\revisioncolor}}
            \hspace{1.9em}
            \subfloat{\includegraphics[width=0.60\columnwidth]{figures/single_query_bar_legend.png}} \hfill%
            \vspace{-1.1em}
            \setcounter{subfigure}{0}

            \subfloat[\revision{\textit{WILDS}}]{\includegraphics[width=0.48\linewidth]{figures/single_query_bar_wilds.png}%
                \label{subfig:single-query-wilds}}%
            \hfil
            \subfloat[\revision{\textit{ImageNet}}]{\includegraphics[width=0.48\linewidth]{figures/single_query_bar_imagenet.png}%
                \label{subfig:single-query-imagenet}}%
        \end{center}
        \vspace{-1.2em}
        \caption{End-to-end individual query execution time based on motivation and related work. The index size for \system is $\sim5\%$ of the original compressed dataset size for both datasets. Note the log scale on the y-axis.}
        \vspace{-0.8em}
        \label{fig:single-query-performance}
    \end{figure*}
}

\newcommand{\queryTimeVsQueryTypeFigure}{
    \begin{figure}[t!]
        \vspace{-1.0em}
        \begin{center}

            \subfloat[\revision{\textit{WILDS}}]{\includegraphics[width=0.5\linewidth]{figures/query_time_vs_query_type_boxplot_wilds.png}%
                \label{subfig:query-time-vs-query-type-wilds}}%
            \hfil
            \subfloat[\revision{\textit{ImageNet}}]{\includegraphics[width=0.5\linewidth]{figures/query_time_vs_query_type_boxplot_imagenet.png}%
                \label{subfig:query-time-vs-query-type-imagenet}}%
        \end{center}
        \vspace{-1.2em}
        \caption{Query time of \system for different query types. Index size for \system: $\sim5\%$ of dataset size.}
        \vspace{-1.8em}
        \label{fig:query-time-vs-query-type}
    \end{figure}
}

\newcommand{\queryTimeVsFractionOfMasksLoadedFigure}{
    \begin{figure}[t!]
        \begin{center}

            \subfloat[\textit{WILDS}, Pearson's $r = 0.99$]{\includegraphics[width=0.48\linewidth]{figures/query_time_vs_fml_scatter_wilds.png}%
                \label{subfig:query-time-vs-fraction-of-masks-loaded-wilds}}%
            \hfil
            \subfloat[\textit{ImageNet}, Pearson's $r = 0.96$]{\includegraphics[width=0.5\linewidth]{figures/query_time_vs_fml_scatter_imagenet.png}%
                \label{subfig:query-time-vs-fraction-of-masks-loaded-imagenet}}%
        \end{center}
        \vspace{-1.2em}
        \caption{Relationship between end-to-end query time and the fraction of masks loaded (FML) for a query.}
        \vspace{-1.8em}
        \label{fig:query-time-vs-fraction-of-masks-loaded}
    \end{figure}
}

\newcommand{\combinedBoundSegmentsFigure}{
    \begin{figure*}[t!]
        \begin{center}
            \subfloat[\textit{WILDS, 88 MB, $(0.6, 1.0)$}]{\includegraphics[width=0.25\linewidth]{figures/object_segment_wilds_threshold_0.6.png}\vspace{-0.8em}}%
            \hfil
            \subfloat[\textit{WILDS, 88 MB, $(0.8, 1.0)$}]{\includegraphics[width=0.25\linewidth]{figures/object_segment_wilds_threshold_0.8.png}\vspace{-0.8em}}%
            \hfil
            \subfloat[\textit{WILDS, 2.2 GB, $(0.6, 1.0)$}]{\includegraphics[width=0.25\linewidth]{figures/object_segment_wilds_threshold_0.6_32_16.png}\vspace{-0.8em}}%
            \hfil
            \subfloat[\textit{WILDS, 2.2 GB, $(0.8, 1.0)$}]{\includegraphics[width=0.25\linewidth]{figures/object_segment_wilds_threshold_0.8_32_16.png}\vspace{-0.8em}}%
            \vspace{-1.1em}
            \subfloat[\textit{ImageNet, 6.5 GB, $(0.6, 1.0)$}]{\includegraphics[width=0.25\linewidth]{figures/object_segment_imagenet_threshold_0.6.png}\vspace{-0.8em}}%
            \hfil
            \subfloat[\textit{ImageNet, 6.5 GB, $(0.8, 1.0)$}]{\includegraphics[width=0.25\linewidth]{figures/object_segment_imagenet_threshold_0.8.png}\vspace{-0.8em}}%
            \hfil
            \subfloat[\textit{ImageNet, 23 GB, $(0.6, 1.0)$}]{\includegraphics[width=0.25\linewidth]{figures/object_segment_imagenet_threshold_0.6_16_14.png}\vspace{-0.8em}}%
            \hfil
            \subfloat[\textit{ImageNet, 23 GB, $(0.8, 1.0)$}]{\includegraphics[width=0.25\linewidth]{figures/object_segment_imagenet_threshold_0.8_16_14.png}\vspace{-0.8em}}%
        \end{center}
        \vspace{-1.2em}

        \caption{Distribution of bounds of $\texttt{CP}(mask, roi, (lv, uv))$ computed by \system. Each subfigure represents the distribution for a combination of $(\text{dataset}, \text{index size}, (lv, uv))$, shown as the title of each. Each vertical segment represents the lower and upper bounds of $\texttt{CP}(mask, roi, (lv, uv))$ for a single mask. For each mask, $roi$ is the foreground object bounding box. We show the distribution of bounds for \num{1000} randomly sampled masks in each subplot. The x-axes represent the masks sorted by their lower bounds. The horizontal dashed lines represent examples of the count threshold $T$. FML is the fraction of masks loaded by \system given a predicate $\texttt{CP}(mask, roi, (lv, uv)) > T$. For each count threshold $T$, FML is equal to the fraction of the vertical segments that intersect with the horizontal dashed line defined by $T$. Note the different scales of the y-axes.}
        \vspace{-1.0em}
        \label{fig:combined-bound-segments}
    \end{figure*}
}

\newcommand{\multiQueryWorkloadFigure}{
    \begin{figure*}[t!]
        \begin{center}
            \captionsetup{font={color=black}}
            \hspace{1em}
            \subfloat{\includegraphics[width=0.6225\columnwidth]{figures/multi_query_workload_cumulative_times_legend.png}}%
            \setcounter{subfigure}{0}
            \hfil
            \hspace{11.50em}
            \subfloat{\includegraphics[width=0.660\columnwidth]{figures/multi_query_workload_speedups_workloads_legend.png}} \hfill%
            \vspace{-1.1em}
            \setcounter{subfigure}{0}

            \subfloat[\revision{\textit{WILDS, Workload 2}}]{\includegraphics[width=0.24\linewidth]{figures/multi_query_workload_cumulative_times_wilds_[0.1, 0.2, 0.3]_0.5_200.png}%
                \label{subfig:multi-query-workload-a}}%
            \hfil
            \subfloat[\revision{\textit{ImageNet, Workload 2}}]{\includegraphics[width=0.24\linewidth]{figures/multi_query_workload_cumulative_times_imagenet_[0.1, 0.2, 0.3]_0.5_200.png}%
                \label{subfig:multi-query-workload-b}}%
            \hfil
            \subfloat[\revision{\textit{WILDS, MS-II vs. MS}}]{\includegraphics[width=0.24\linewidth]{figures/multi_query_workload_speedups_workloads_wilds.png}%
                \label{subfig:multi-query-workload-c}}%
            \hfil
            \subfloat[\revision{\textit{ImageNet, MS-II vs. MS}}]{\includegraphics[width=0.24\linewidth]{figures/multi_query_workload_speedups_workloads_imagenet.png}%
                \label{subfig:multi-query-workload-c}}%
        \end{center}
        \vspace{-1.0em}
        \caption{Cumulative total time, incl. index building time and query time, for multi-query workloads. MS-II and MS refer to \system w/ and w/o incremental indexing, respectively. (a) and (b) show the total time for MS, MS-II, and NumPy for Workload 2; (c) and (d) show the ratio of the cumulative total time of MS-II to that of MS for all workloads. The index size for MS is $\sim5\%$ of the corresponding dataset. MS-II builds the index incrementally using the same index configuration as MS.}
        \label{fig:multi-query-workload}
    \end{figure*}
}

\vspace{-2em}
\section{Introduction} \label{sec:intro}

Masking is a way to highlight or isolate certain parts of an image based on desired properties for further processing or analysis. 
Machine learning tasks over image databases often involve generating and using masks, such as image segmentation masks~\cite{yolo} and model saliency maps~\cite{gradcam2017}. 
These masks are crucial for a wide range of applications, from model explanation~\cite{gradcam2017, degrave2021ai} to real-world analysis~\cite{datafromsky-traffic}. 
For example, practitioners developing image classification models can generate model saliency maps to understand which pixels contribute the most to the model's predictions. 

\explainMask

\textit{Consider a scenario further discussed in~\cref{scenarios}, Alice, a data engineer, uses the \textit{iWildCam} dataset~\cite{beery2020iwildcam} for developing a wild animal image classification model. 
Facing validation accuracy issues, she computes saliency maps~\cite{gradcam2017} and YOLO-generated bounding boxes~\cite{yolo} for the misclassified images, an example of which is shown in~\cref{fig:explainMask}. 
In the saliency map, the red pixels indicate higher importance for the model's prediction, and the blue pixels indicate lower importance. 
She finds that the model focuses on the background pixels, notably outside the ground-truth object bounding boxes, rather than the animals, leading to misclassifications when background conditions change. 
To correct the model's focus, Alice wishes to augment the dataset and retrain the model to ensure that it relies on relevant features to make predictions. She first retrieves a group of images where the model focuses on the area outside the intended object bounding box. 
She then augments the dataset with these images modified by randomizing pixels outside object bounding boxes while leaving the original labels unchanged and retrains the model with the augmented dataset. 
The retrained model will have improved accuracy on the validation set by focusing on relevant pixels~\cite{teso2023leveraging}.}

As the scenario shows, the ability to retrieve images and masks based on the properties of the latter is valuable to machine learning workflows, and the diverse applications of image masks stress this need for ML practitioners. 
However, efficient execution of these queries suffers from insufficient systems support~\cite{hong2020human}. 

We recently developed \system~\cite{he2024masksearch}, a system that addresses this challenge by accelerating queries over databases of image masks.
\system's contributions include formalizing a class of image and mask retrieval queries with support for aggregations and top-$k$ computations, introducing a novel indexing technique over masks and an efficient execution framework, and implementing a prototype that significantly outperforms existing solutions in query execution efficiency for both individual and multi-query workloads that simulate machine learning workflows.

In this demonstration, we introduce a graphical user interface (GUI) for \system (\cref{sec:ui}), which enables users to execute queries without writing SQL and conveniently displays images, masks, and bounding boxes. 
We also illustrate \system's utility across multiple scenarios (\cref{scenarios}) in addition to the aforementioned scenario:

\begin{itemize}[itemsep=1pt, topsep=1pt, leftmargin=10pt]
    \item \textit{Scenario 2} demonstrates how \system can assist in identifying adversarial attacks. We show the ability of \system to retrieve maliciously attacked images in a dataset by calculating the dispersion of model saliency, relieving the effort required for finding attacks unrecognizable to human eyes.

    \item \textit{Scenario 3} demonstrates how \system helps in investigating discrepancies between model saliency and human attention.
\end{itemize}

Overall, this demonstration will enable conference attendees to experiment with \system hands-on, appreciate the convenience and performance of the system, as well as better understand its inner workings.

\vspace{-1em}
\section{System Overview}

In this section, we summarize the \system system, as detailed in~\cite{he2024masksearch}, focusing on its principal capabilities and features.

\noindent \textbf{Data Model.} An image mask is a 2D array of pixel values represented by floating-point numbers within the range of $[0, 1)$. \system supports queries over a database of masks by exposing those masks through a conceptually relational view with one attribute holding the mask data and the other attributes capturing the mask metadata.%

\small
\begin{verbatim}
MasksDatabaseView (
  mask_id INTEGER PRIMARY KEY,
  image_id INTEGER,  // Image from which mask was derived
  model_id INTEGER,  // Model that generated the mask
  mask_type INTEGER, // Type of mask (e.g., saliency map)
  mask REAL[][]);
\end{verbatim}
\normalsize

\noindent \textbf{Region of Interest (ROI).} An ROI is defined by a bounding box that specifies the area of interest within a mask. It is not included in \texttt{MasksDatabaseView} since it is query-dependant and may
be computed on the fly (e.g., object detector applied to the image). %

\noindent \textbf{CP Function.} CP stands for ``Count Pixels''. \small$\texttt{CP}(mask, roi, (lv, uv))$ \normalsize counts the number of pixels within the ROI in the mask whose values fall within the specified value range $[lv, uv)$. Users can use multiple \small \texttt{CP} \normalsize functions and apply arithmetic operations in queries.

\system supports various query types, including filter queries, top-k queries, and aggregation queries, as detailed below.

\noindent \textbf{Filter Query.} This query type retrieves masks based on filter conditions on \small $\texttt{CP}(mask, roi, (lv, uv))$\normalsize. The filter condition is defined by a threshold T and an inequality symbol.

\small
\begin{verbatim}
SELECT mask_id FROM MasksDatabaseView
WHERE CP(mask, roi, (lv, uv)) < T;
\end{verbatim}
\normalsize

\noindent \textbf{Top-K Query.} This query type retrieves the top-$k$ masks ranked by  \small$\texttt{CP}(mask, roi, (lv, uv))$\normalsize. The ranking order can be ascending (\texttt{ASC}) or descending (\texttt{DESC}). 
\small
\begin{verbatim}
SELECT mask_id FROM MasksDatabaseView
ORDER BY CP(mask, roi, (lv, uv)) DESC LIMIT K;
\end{verbatim}
\normalsize

\noindent \textbf{Aggregation Query.} \system supports both scalar aggregation and mask aggregation. For scalar aggregation, the user can aggregate the outputs of \texttt{CP} functions through the \texttt{SCALAR\_AGG} function. 
\system supports aggregation functions like \texttt{SUM}, \texttt{AVG}, \texttt{MIN}, and \texttt{MAX}. 
Mask aggregation facilitates the combination or comparison of information across multiple masks (of the same image), treating aggregated masks as new queryable entities. 
The user needs to define a function \texttt{MASK\_AGG} that takes in a list of masks and returns an aggregated mask: \texttt{MASK\_AGG}$\rightarrow$ \texttt{REAL[][]}, where \texttt{MASK\_AGG} can be any function \small$f(m_1, m_2, ..., m_n)$\normalsize, where \small$m_i$\normalsize represents a mask. 
For example, \small$\texttt{intersect}(m_1 > 0.8, ..., m_n > 0.8)$\normalsize outputs the intersection of the masks \small$m_1, ..., m_n$\normalsize thresholded by 0.8.

\small
\begin{verbatim}
SELECT image_id FROM MasksDatabaseView
WHERE mask_type IN (1, 2, ..., n)
GROUP BY image_id ORDER BY CP(MASK_AGG(mask), roi, (lv, uv));
\end{verbatim}
\normalsize

To efficiently support these queries, \system introduces two key components: the Cumulative Histogram Index (CHI) and a filter-verification query execution framework. 
CHI is a novel indexing technique that stores pixel counts for different key combinations of spatial locations and pixel values, which enables the efficient derivation of upper and lower bounds for pixel counts of arbitrary ROIs and pixel value ranges specified by the user at query time. 
The filter-verification framework leverages CHI to compute bounds to determine which masks can be added directly to the result set or pruned without loading them from disk to memory and which require further verification by loading them from disk and applying the predicate. 
As is further illustrated in~\cite{he2024masksearch}, this approach significantly reduces disk I/O which is the bottleneck for query execution. 
In this demonstration, attendees will be able to experience both \system's ease-of-use and query performance, as well as explore how \system executes queries.

\vspace{-1em}
\section{\system Interface} \label{sec:ui}
\scenarioOneFigure

\noindent This section describes \system’s interface (\cref{fig:scenarioOne}).

In the Data Preparation phase, \system’s interface allows users to load and specify their models, datasets, and masks. 
\system can compute a variety of masks, or the users can provide the masks. 
This process is followed by the automatic calculation and display of the model's accuracy and a confusion matrix where each clickable cell represents the images whose ground truth label and predicted label are the corresponding row and column of the cell, respectively. 
For example, Cell (146, 17) represents images of class 146 that were classified as class 17. 
Before clicking on a cell, users can optionally load regions of interest for the images, e.g., object bounding boxes computed by an off-the-shelf model. 
As illustrated in Step 1 in~\cref{fig:scenarioOne}, this functionality allows for detailed visualization of the images from the selected cell (146, 17) with their corresponding masks (saliency maps in this example) and, optionally, the rectangle boxes representing the ROI. 
The interface also provides a figure illustrating how \system builds its indexes (CHI) and how CHI is used to accelerate query execution. 
Due to space constraints, the initial data loading, confusion matrix, and the illustrative figure for CHI are not presented in~\cref{fig:scenarioOne}.

\noindent \textbf{Input Section.} The Input Section
is demonstrated on the left of Steps 2 and 3 in Figure~\ref{fig:scenarioOne}. 
It simplifies the creation and manipulation of search queries by providing a form that guides users through specifying their query, including defining an optional ROI (ROI is the full mask by default), upper and lower bounds of the pixel value range, and choosing between different queries such as Top-K Query, Filter Query, and Aggregation Query. 
The interface enables the ROI definition through two approaches: (1) mask-dependent ROIs provided by the user, such as object bounding boxes generated by an off-the-shelf model, as introduced in the Data Preparation phase, or (2) a constant ROI across masks: a rectangle drawn by the user on an image. 
Based on the aforementioned user-specified parameters, the interface generates an SQL query shown in the "Query Command" window, which allows the users to inspect the formalized query and, if necessary, directly modify the SQL query for their search. 
After a query is executed, clicking "Execution Detail" triggers the interface to show the distribution of the lower and upper bounds for \small\texttt{CP$(mask, roi, (lv, uv))$}\normalsize computed by \system for the users to understand how \system reduces the number of masks that must be loaded from disk during query execution while guaranteeing the correct result.

\noindent \textbf{Query Result Section.} 
The Query Result Section, presented on the right of Steps 2 and 3 in~\cref{fig:scenarioOne}, displays the query results as a combination of images and their corresponding masks, dependent on the specific scenario. 
For example, in Step 2 of~\cref{fig:scenarioOne}, the returned images are overlaid with their saliency maps and the object bounding boxes. 
The UI also offers users the ability to click and zoom in on the query results in a popup window. 

\noindent \textbf{Dataset Augmentation.} 
To extend the capabilities of \system's interface for machine learning workflows, this demonstration also incorporates a dataset augmentation feature, which is further described in~\cref{sec:demo}.

\vspace{-1.1em}
\section{Demonstration Scenarios} \label{sec:demo}

Our demonstration will walk through a series of scenarios that show \system's utility in real-world machine learning workflows:

\label{scenarios}\textbf{Scenario 1: Debugging Image Classification Models}~\cite{plumb2021finding, teso2023leveraging}, illustrated in~\cref{fig:scenarioOne}. 
Recall the scenario mentioned in~\cref{sec:intro}. Alice noticed that the model learned to rely on the presence of confounding factors in the background to classify the animals, as shown in Step 1 in~\cref{fig:scenarioOne}. To mitigate the model's reliance on confounding factors, Alice can first use a Top-K query to retrieve the images with the least number of high-value pixels in the ROI (object bounding box generated by YOLO~\cite{yolo}) normalized by the area of the ROI, as shown in Step 2 in~\cref{fig:scenarioOne}. Another option is to use a Filter query to retrieve all the images for which the normalized number of high-value pixels in the ROI is below a threshold. She can then augment her training set by randomizing the pixels outside the ROI in the retrieved images with the original labels, as shown in Step 3 in~\cref{fig:scenarioOne}, and retrain her model on the augmented training set, which guides the model to classify the animals without relying on the randomized background pixels.

In this scenario, we demonstrate \system’s ability to execute Top-K and Filter queries efficiently. 
On an AWS EC2 p3.2xlarge instance which has an Intel Xeon E5-2686 v4 processor with 8 vCPUs and 61 GiB of memory, 
and EBS gp3 volumes provisioned with 3000 IOPS and 125 MiB/s throughput for disk storage, without \system, the median execution times of 5 Filter queries and 5 Top-K queries (OS page cache cleared before each run) on $\num{22275}$ images (with their model saliency masks) from the \textit{iWildCam} dataset~\cite{beery2020iwildcam} are both around 100 seconds. 
In contrast, it takes \system less than a second to execute the same queries (OS page cache cleared before each run), which is a $100\times$ speedup.

The conference attendees will interact with \system using the interface shown in~\cref{fig:scenarioOne}. They will be able to explore misclassified images and
execute Top-K and Filter queries (we will pre-populate the fields and the attendees will
be able to change the values).
After clicking "Start Query", the attendees will see the images overlaid by their corresponding saliency maps returned by the query. Attendees will also be able to click the "Start Augment" button to augment those images, and the result will be shown on the interface. Finally, we will
provide an additional tab showing the details of the CHI and how different image masks
were effectively filtered during query execution.

\textbf{Scenario 2: Identifying Adversarial Attacks}~\cite{ye2022detection}. 
Claudia is an ML engineer who develops and maintains an image classification model that performs with high accuracy in production. 
During a routine check, she discovers that there is a significant drop in the prediction accuracy. 
Claudia examines the misclassified images manually and they look normal. 
However, after computing the model saliency maps for those images, she notices that the model's attention is diffused across irrelevant regions
similar to the example shown in ~\cref{fig:scenarioTwo} (b).
Hence, she starts to suspect the misclassification may be due to malicious modifications that mislead the model to focus on irrelevant pixels. 
She wishes to retrieve the saliency maps that contain the most mid-value pixels, which indicates diffused model attention. 
With \system, she specifies the ROI as the full mask and issues a Top-K query. 
An example query she might use is,

\small
\begin{verbatim}
SELECT mask_id FROM MasksDatabaseView
ORDER BY CP(mask, full_img, (0.2, 0.6)) DESC LIMIT 25;
\end{verbatim}
\normalsize

By examining the returned masks (and their corresponding images), Claudia could better understand whether (and to what extent) the images were maliciously modified and improve the model’s resilience to such malicious modifications.

\scenarioTwoFigure

The conference attendees will be able to traverse through the scenario with the same interface shown in~\cref{fig:scenarioOne}. 
They will first see both attacked and unattacked images and their corresponding saliency maps shown side-by-side to explore different patterns, e.g., focused attention vs. diffused attention, between the two categories;
Based on the observation of which range the majority of diffused attention pixel values fall within, they can establish custom upper and lower bounds in Top-K query to obtain masks with the most (or least) diffused attention.
Attendees will be able to examine the model saliency maps overlaid on the returned images.

\HAFigure

\textbf{Scenario 3:  Investigating discrepancies between model saliency and human attention}~\cite{das2017human}. 
This scenario demonstrates \system's ability to perform aggregation queries efficiently. 
Fine-grained image classification requires identifying local and discriminate regions that correspond to subtle visual traits. Exploiting human attention can rectify models that deviate from critical traits for making correct predictions~\cite{rong2021human}. 
An example is illustrated in~\cref{fig:HA}. 
Imagine a scenario in which a human-centered AI researcher, Bob, wants to investigate whether a fine-grained classification model is looking at the same region as humans to make the prediction. He first thresholds the saliency maps and human attention maps to binary masks (pixels > threshold becomes 1; otherwise 0) to reduce noises in the masks. With \system, he can then efficiently retrieve the images where the attention of the model and human experts has the lowest degree of alignment by aggregating the human attention and model saliency masks (group by \texttt{image\_id}) and computing the Intersection over Union (IoU). An example query he might use is shown below:

\small
\begin{verbatim}
SELECT image_id, 
    CP(intersect(mask > 0.8), roi, (lv, uv)) 
        / CP(union(mask > 0.8), roi, (lv, uv)) as iou
FROM MasksDatabaseView WHERE mask_type IN (1, 2)
GROUP BY image_id ORDER BY iou ASC LIMIT 25;
\end{verbatim}
\normalsize

In this scenario, the conference attendees will be guided to execute aggregation queries on the given human attention map and model saliency map with \system. They need to define a value T for thresholding the two masks and either start a Filter query or a Top-K query following the same input procedure described in Scenario 1, except that the ROI is set to the whole image. The query will return a list of images where the human attention map and model saliency map have the lowest IoU. Attendees will see the two masks of those images presented side-by-side on the GUI.

\section{Related work}

Although prior work has proposed systems that support queries over image databases~\cite{beaver2010finding, vdms2021remis, qbic1995flickner}, these methods are not optimized for \system's target queries that retrieve images and masks based on mask properties. 
\system falls into the group of systems that support ML model inspection, explanation, and debugging~\cite{sellam2019deepbase, mehta2020toward, deepeverest2021he}, among which DeepEverest~\cite{deepeverest2021he} is most relevant to \system. 
DeepEverest helps practitioners better understand neural network behavior by supporting the efficient querying of input examples based on neural representations. 
While \system also focuses on efficiently retrieving examples, it targets a fundamentally different class of queries based on mask properties.

\end{sloppypar} %

\bibliographystyle{ACM-Reference-Format}
\bibliography{sample}


\begin{thebibliography}{20}

%%% the Web address.

\ifx \showCODEN    \undefined \def \showCODEN     #1{\unskip}     \fi
\ifx \showDOI      \undefined \def \showDOI       #1{#1}\fi
\ifx \showISBNx    \undefined \def \showISBNx     #1{\unskip}     \fi
\ifx \showISBNxiii \undefined \def \showISBNxiii  #1{\unskip}     \fi
\ifx \showISSN     \undefined \def \showISSN      #1{\unskip}     \fi
\ifx \showLCCN     \undefined \def \showLCCN      #1{\unskip}     \fi
\ifx \shownote     \undefined \def \shownote      #1{#1}          \fi
\ifx \showarticletitle \undefined \def \showarticletitle #1{#1}   \fi
\ifx \showURL      \undefined \def \showURL       {\relax}        \fi
% invisible to TeX
\providecommand\bibfield[2]{#2}
\providecommand\bibinfo[2]{#2}
\providecommand\natexlab[1]{#1}
\providecommand\showeprint[2][]{arXiv:#2}

\bibitem[\protect\citeauthoryear{DataFromSky}{DataFromSky}{2023}]%
        {datafromsky-traffic}
\bibfield{author}{\bibinfo{person}{DataFromSky}.} \bibinfo{year}{2023}\natexlab{}.
\newblock \bibinfo{title}{Traffic Monitoring - DataFromSky}.
\newblock
\newblock
\urldef\tempurl%
\url{https://datafromsky.com/traffic-monitoring/}
\showURL{%
\tempurl}


\bibitem[\protect\citeauthoryear{el~al.}{el~al.}{2017}]%
        {das2017human}
\bibfield{author}{\bibinfo{person}{Das el al.}} \bibinfo{year}{2017}\natexlab{}.
\newblock \showarticletitle{Human attention in visual question answering: Do humans and deep networks look at the same regions?}
\newblock \bibinfo{journal}{\emph{Computer Vision and Image Understanding}} (\bibinfo{year}{2017}).
\newblock


\bibitem[\protect\citeauthoryear{et~al.}{et~al.}{2010}]%
        {beaver2010finding}
\bibfield{author}{\bibinfo{person}{Beaver et al.}} \bibinfo{year}{2010}\natexlab{}.
\newblock \showarticletitle{Finding a Needle in Haystack: Facebook's Photo Storage}. In \bibinfo{booktitle}{\emph{OSDI}}, Vol.~\bibinfo{volume}{10}. \bibinfo{pages}{1--8}.
\newblock


\bibitem[\protect\citeauthoryear{et~al.}{et~al.}{2020a}]%
        {beery2020iwildcam}
\bibfield{author}{\bibinfo{person}{Beery et al.}} \bibinfo{year}{2020}\natexlab{a}.
\newblock \showarticletitle{The iWildCam 2020 Competition Dataset}.
\newblock \bibinfo{journal}{\emph{arXiv preprint arXiv:2004.10340}} (\bibinfo{year}{2020}).
\newblock


\bibitem[\protect\citeauthoryear{et~al.}{et~al.}{2009}]%
        {deng2009imagenet}
\bibfield{author}{\bibinfo{person}{Deng et al.}} \bibinfo{year}{2009}\natexlab{}.
\newblock \showarticletitle{Imagenet: A large-scale hierarchical image database}. In \bibinfo{booktitle}{\emph{CVPR}}. \bibinfo{pages}{248--255}.
\newblock


\bibitem[\protect\citeauthoryear{et~al.}{et~al.}{2021a}]%
        {degrave2021ai}
\bibfield{author}{\bibinfo{person}{DeGrave et al.}} \bibinfo{year}{2021}\natexlab{a}.
\newblock \showarticletitle{AI for radiographic COVID-19 detection selects shortcuts over signal}.
\newblock \bibinfo{journal}{\emph{Nature Machine Intelligence}} \bibinfo{volume}{3}, \bibinfo{number}{7} (\bibinfo{year}{2021}), \bibinfo{pages}{610--619}.
\newblock


\bibitem[\protect\citeauthoryear{et~al.}{et~al.}{1995}]%
        {qbic1995flickner}
\bibfield{author}{\bibinfo{person}{Flickner et al.}} \bibinfo{year}{1995}\natexlab{}.
\newblock \showarticletitle{Query by image and video content: the QBIC system}.
\newblock \bibinfo{journal}{\emph{Computer}} \bibinfo{volume}{28}, \bibinfo{number}{9} (\bibinfo{year}{1995}), \bibinfo{pages}{23--32}.
\newblock


\bibitem[\protect\citeauthoryear{et~al.}{et~al.}{2020b}]%
        {hong2020human}
\bibfield{author}{\bibinfo{person}{Hong et al.}} \bibinfo{year}{2020}\natexlab{b}.
\newblock \showarticletitle{Human factors in model interpretability: Industry practices, challenges, and needs}.
\newblock \bibinfo{journal}{\emph{PACM HCI}} \bibinfo{volume}{4}, \bibinfo{number}{CSCW1} (\bibinfo{year}{2020}), \bibinfo{pages}{1--26}.
\newblock


\bibitem[\protect\citeauthoryear{et~al.}{et~al.}{2021b}]%
        {deepeverest2021he}
\bibfield{author}{\bibinfo{person}{He et al.}} \bibinfo{year}{2021}\natexlab{b}.
\newblock \showarticletitle{DeepEverest: Accelerating Declarative Top-K Queries for Deep Neural Network Interpretation}.
\newblock \bibinfo{journal}{\emph{Proc. VLDB Endow.}} \bibinfo{volume}{15}, \bibinfo{number}{1} (\bibinfo{year}{2021}), \bibinfo{pages}{98–111}.
\newblock
\showISSN{2150-8097}


\bibitem[\protect\citeauthoryear{et~al.}{et~al.}{2023a}]%
        {he2024masksearch}
\bibfield{author}{\bibinfo{person}{He et al.}} \bibinfo{year}{2023}\natexlab{a}.
\newblock \showarticletitle{MaskSearch: Querying Image Masks at Scale}.
\newblock \bibinfo{journal}{\emph{arXiv preprint arXiv:2305.02375}} (\bibinfo{year}{2023}).
\newblock


\bibitem[\protect\citeauthoryear{et~al.}{et~al.}{2020c}]%
        {mehta2020toward}
\bibfield{author}{\bibinfo{person}{Mehta et al.}} \bibinfo{year}{2020}\natexlab{c}.
\newblock \showarticletitle{Toward Sampling for Deep Learning Model Diagnosis}. In \bibinfo{booktitle}{\emph{ICDE}}. IEEE, \bibinfo{pages}{1910--1913}.
\newblock


\bibitem[\protect\citeauthoryear{et~al.}{et~al.}{2021c}]%
        {plumb2021finding}
\bibfield{author}{\bibinfo{person}{Plumb et al.}} \bibinfo{year}{2021}\natexlab{c}.
\newblock \showarticletitle{Finding and fixing spurious patterns with explanations}.
\newblock \bibinfo{journal}{\emph{arXiv preprint arXiv:2106.02112}} (\bibinfo{year}{2021}).
\newblock


\bibitem[\protect\citeauthoryear{et~al.}{et~al.}{2016}]%
        {yolo}
\bibfield{author}{\bibinfo{person}{Redmon et al.}} \bibinfo{year}{2016}\natexlab{}.
\newblock \showarticletitle{You only look once: Unified, real-time object detection}. In \bibinfo{booktitle}{\emph{CVPR}}. \bibinfo{pages}{779--788}.
\newblock


\bibitem[\protect\citeauthoryear{et~al.}{et~al.}{2021d}]%
        {rong2021human}
\bibfield{author}{\bibinfo{person}{Rong et al.}} \bibinfo{year}{2021}\natexlab{d}.
\newblock \showarticletitle{Human attention in fine-grained classification}.
\newblock \bibinfo{journal}{\emph{arXiv preprint arXiv:2111.01628}} (\bibinfo{year}{2021}).
\newblock


\bibitem[\protect\citeauthoryear{et~al.}{et~al.}{2021e}]%
        {vdms2021remis}
\bibfield{author}{\bibinfo{person}{Remis et al.}} \bibinfo{year}{2021}\natexlab{e}.
\newblock \showarticletitle{Using VDMS to Index and Search 100M Images}.
\newblock \bibinfo{journal}{\emph{Proc. VLDB Endow.}} \bibinfo{volume}{14}, \bibinfo{number}{12} (\bibinfo{year}{2021}), \bibinfo{pages}{3240–3252}.
\newblock
\showISSN{2150-8097}


\bibitem[\protect\citeauthoryear{et~al.}{et~al.}{2017}]%
        {gradcam2017}
\bibfield{author}{\bibinfo{person}{Selvaraju et al.}} \bibinfo{year}{2017}\natexlab{}.
\newblock \showarticletitle{Grad-CAM: Visual Explanations from Deep Networks via Gradient-Based Localization}. In \bibinfo{booktitle}{\emph{ICCV}}. \bibinfo{pages}{618--626}.
\newblock


\bibitem[\protect\citeauthoryear{et~al.}{et~al.}{2019}]%
        {sellam2019deepbase}
\bibfield{author}{\bibinfo{person}{Sellam et al.}} \bibinfo{year}{2019}\natexlab{}.
\newblock \showarticletitle{Deepbase: Deep inspection of neural networks}. In \bibinfo{booktitle}{\emph{SIGMOD}}. \bibinfo{pages}{1117--1134}.
\newblock


\bibitem[\protect\citeauthoryear{et~al.}{et~al.}{2023b}]%
        {teso2023leveraging}
\bibfield{author}{\bibinfo{person}{Teso et al.}} \bibinfo{year}{2023}\natexlab{b}.
\newblock \showarticletitle{Leveraging explanations in interactive machine learning: An overview}.
\newblock \bibinfo{journal}{\emph{Frontiers in Artificial Intelligence}}  \bibinfo{volume}{6} (\bibinfo{year}{2023}).
\newblock


\bibitem[\protect\citeauthoryear{et~al.}{et~al.}{2011}]%
        {WahCUB_200_2011}
\bibfield{author}{\bibinfo{person}{Wah et al.}} \bibinfo{year}{2011}\natexlab{}.
\newblock \bibinfo{booktitle}{\emph{Caltech-UCSD Birds-200-2011 Dataset}}.
\newblock \bibinfo{type}{{T}echnical {R}eport} CNS-TR-2011-001. \bibinfo{institution}{California Institute of Technology}.
\newblock


\bibitem[\protect\citeauthoryear{et~al.}{et~al.}{2022}]%
        {ye2022detection}
\bibfield{author}{\bibinfo{person}{Ye et al.}} \bibinfo{year}{2022}\natexlab{}.
\newblock \showarticletitle{Detection defense against adversarial attacks with saliency map}.
\newblock \bibinfo{journal}{\emph{International Journal of Intelligent Systems}} \bibinfo{volume}{37}, \bibinfo{number}{12} (\bibinfo{year}{2022}).
\newblock


\end{thebibliography}

\end{document}